\documentclass[12pt]{article}
\usepackage{graphicx,amsmath,amssymb}
\usepackage{indentfirst}
\usepackage[usenames]{color}
\usepackage[colorlinks=true, urlcolor=navyblue, linkcolor=navyblue, citecolor=navyblue]{hyperref}
\usepackage{epstopdf}
\usepackage{enumerate}
\usepackage{appendix}
\usepackage{lineno}
\usepackage{setspace}

\definecolor{navyblue}{rgb}{0,0.08,0.45}
\definecolor{darkred}{rgb}{0.7,0.0,0.0}
\definecolor{darkgreen}{rgb}{0,0.6,0.2}

\newcommand{\beq}{\begin{equation}}
\newcommand{\enq}{\end{equation}}
\newcommand{\beqa}{\begin{eqnarray}}
\newcommand{\beqast}{\begin{eqnarray*}}
\newcommand{\enqa}{\end{eqnarray}}
\newcommand{\enqast}{\end{eqnarray*}}

\newcommand{\bec}{\begin{center}}
\newcommand{\enc}{\end{center}}
\newcommand{\beqo}{\begin{quote}}
\newcommand{\enqo}{\end{quote}}
\newcommand{\bem}{\begin{minipage}}
\newcommand{\enm}{\end{minipage}}

\begin{document}

\vspace{15pt}

\begin{center}
{\large Supersymmetric Meson-Baryon Properties of QCD 
from Light-Front Holography and Superconformal Algebra}

\vspace{10pt}


\end{center}

\vspace{15pt}

\centerline{Stanley J. Brodsky}

\vspace{3pt}

\centerline {\it SLAC National Accelerator Laboratory, Stanford University~\footnote{{Invited talk presented at Light Cone 2016,  
September 5-8, 2016, Centro de Fisca Teo'rica de Particulas, CFTP, IST University of Lisbon, Portugal\\
\href{mailto:sjbth@slac.stanford.edu}{\tt
\hspace{12pt} sjbth@slac.stanford.edu}}}}

\vspace{20pt}

\begin{abstract}

A remarkable feature of QCD is that the mass scale $\kappa$ which controls color confinement  and light-quark  hadron mass scales does not appear explicitly in the QCD Lagrangian. However,  de Alfaro, Fubini, and Furlan have shown that a mass scale can appear in the equations of motion without affecting the conformal invariance of the action if one adds a term to the Hamiltonian proportional to the dilatation operator or the special conformal operator.    If one applies the same procedure to the light-front Hamiltonian, it leads uniquely to a confinement potential $\kappa^4 \zeta^2$ for mesons, where $\zeta^2$ is the LF radial variable conjugate to  the $q \bar q$ invariant mass.  The same result, including spin terms, is obtained using  light-front holography  -- the duality between the front form and AdS$_5$, the space of isometries of the conformal group -- if one  
modifies the action of AdS$_5$ by the dilaton $e^{\kappa^2 z^2}$ in the fifth dimension $z$.   When one generalizes this procedure using superconformal algebra, the resulting light-front eigensolutions predict a unified Regge spectroscopy of meson, baryon, and tetraquarks, including remarkable supersymmetric relations between the masses of mesons and baryons of the same parity.  
One also predicts observables such as hadron structure functions, transverse momentum distributions, and the distribution amplitudes defined from the hadronic light-front wavefunctions.   
The mass scale $\kappa$ underlying confinement and hadron masses  can be connected to the parameter   $\Lambda_{\overline {MS}}$ in the QCD running coupling by matching the nonperturbative dynamics to the perturbative QCD  regime. The result is an effective coupling $\alpha_s(Q^2)$  defined at all momenta.   The
matching of the high and low momentum transfer regimes determines a scale $Q_0$ which  sets the interface between perturbative and nonperturbative hadron dynamics.  
The use of $Q_0$ to  resolve  the factorization scale uncertainty for structure functions and distribution amplitudes,  in combination with the principle of maximal conformality (PMC)  for  setting the  renormalization scales,  can 
greatly improve the precision of perturbative QCD predictions for collider phenomenology.  The absence of vacuum excitations of the causal, frame-independent  front-form vacuum has important consequences  for the cosmological constant.   I also discuss evidence that the antishadowing of nuclear structure functions is non-universal; {\it i.e.},  flavor dependent, and why shadowing and antishadowing phenomena may be incompatible with sum rules for nuclear parton distribution functions.     \end{abstract}

\tableofcontents

\newpage

\section{Introduction}

Light-front wavefunctions provide a direct link between the QCD Lagrangian and hadron structure.   
The LFWFs are the eigenstates of the Front-Form Hamiltonian $P^- = i {d\over d\tau}$ , the LF time $\tau = x^+ = t + z/c$ evolution operator.
When one makes a measurement of a hadron, such as in deep inelastic lepton-proton scattering $\ell p \to \ell^\prime X$, the hadron is observed
along a light-front (LF) --  in analogy to a flash photograph -- not at a fixed time $t$. This is the underlying principle of the ``front form" discussed  by Dirac~\cite{Dirac:1949cp}.

In the  case of QCD, the eigenvalues of the LF invariant Hamiltonian $ H_{LF} =  P^+ P^- -{\vec P}^2_\perp$, where $P^+=P^0+P^z$ and $\vec P_\perp$ are kinematical, are the squares of the hadron masses $M^2_H$: 
$H_{LF}|\Psi_H>  = M^2_H |\Psi_H>$~\cite{Brodsky:1997de}.  The eigensolutions of $H_{LF} $ provide the $n$-particle hadronic LF Fock state wavefunctions (LFWFs)
 $ \psi^H_n(x_i, \vec k_{\perp i },\lambda_i)= <n| \Psi_H> $, the projection on the free Fock basis.  The  LF Hamiltonian,  can be derived directly from the  QCD Lagrangian.  
The constituents' physical momenta are 
$p^+_i = x_i P^+$, and  $\vec p_{\perp i } =  x_i  {\vec P}_\perp +  \vec k_{\perp i }$,  and the $\lambda_i$ label the  spin projections $S^z_i$.
The LFWFs are Poincare' invariant: they are independent of $P^+$ and $P_\perp$ and are thus independent of the motion of the observer.
The elastic and transition form factors of hadrons, weak-decay amplitudes and distribution amplitudes are overlaps of LFWFs; structure functions, transverse momentum distributions
and other inclusive observables 
are constructed from the squares of the LFWFs.   
The calculation of deeply virtual Compton scattering is given in ref. \cite{Brodsky:2000xy}.
Since the LFWFs are independent of the hadron's momentum, there is no length contraction. 
The absence of length contraction of a photographed object was first noted by Terrell~\cite{Terrell:1959zz} and Penrose~\cite{Penrose:1959vz}. 
One measures the  same structure function in an electron-ion collider as in an electron-scattering experiment where the target hadron is at rest.

Light-front quantization thus provides a physical, frame-independent formalism for hadron dynamics and structure.   The LFWFs play the same role as the Schrodinger wavefunctions which encode the structure of atoms in QED.
One cannot compute form factors of hadrons or other current matrixelements of hadrons from overlap of the usual ``instant" form wavefunctions since one must also include contributions  where the photon interacts with connected but acausal vacuum-induced currents.
One can show that the anomalous gravitomagnetic moment $B(q^2=0)$ vanishes identically for any LF Fock state~\cite{Brodsky:2000ii}, in agreement with a theorem ~\cite{Kobzarev:1962wt,Teryaev:1999su} which follows from the equivalence theorem of gravity.  One can derive $H_{LF}$ directly from the QCD Lagrangian and avoid ghosts and longitudinal  gluonic degrees of freedom by choosing to work in the light-cone gauge  $A^+ =0$.  
The quark masses appear in the LF kinetic energy as $\sum_i {m^2\over x_i}$. This can be derived from the Higgs theory quantized using LF dynamics. 
The confined quark field $\psi_q$ couples to the background Higgs field  $g_{\overline \Psi_q } <H >  \Psi_q$ via its Yukawa  scalar matrix element  coupling  
$g_q <H > \bar u(p) 1 u(p) = m_q \times {m_q \over x} = {m^2\over x}.$

PQCD factorization theorems and  the DGLAP  \cite{Gribov:1972ri,Altarelli:1977zs,Dokshitzer:1977sg} and ERBL \cite{Lepage:1979zb,Lepage:1980fj,Efremov:1979qk,Efremov:1978rn} evolution equations can also be derived using the light-front Hamiltonian formalism~\cite{Lepage:1980fj}.  In the case of an electron-ion collider, one can represent the cross section for $e-p$ colisions as a convolution of the hadron and virtual photon structure functions times the subprocess cross-section in analogy to hadron-hadron colisions.   This nonstandard description of $\gamma^* p \to X$ reactions  gives new insights into electroproduction physics -- physics not apparent   in the usual infinite-momentum frame description, such as the dynamics of heavy quark-pair production.  
Intrinsic heavy quarks at high $x$  also play an important role~\cite{Brodsky:2015uwa}.

The LF Heisenberg equation can in principle be solved numerically by matrix diagonalization  using the ``Discretized Light-Cone  Quantization" (DLCQ)~\cite{Pauli:1985pv} method.  Anti-periodic boundary conditions in 
$x^-$ render the $k^+$ momenta  discrete  as well as  limiting the size of the Fock basis.   In fact, one can easily solve $1+1 $ quantum field theories such as QCD$(1+1)$~\cite{Hornbostel:1988fb} for any number of colors, flavors and quark masses using DLCQ. 
Unlike lattice gauge theory, the nonpertubative DLCQ analysis is in Minkowski space, is frame-independent, and is free of fermion-doubling problems.   
AdS/QCD, based on the  AdS$_5$ representation of the conformal group in five dimensions, maps to physical 3+1 space-time at fixed LF time;  this correspondence, ``light-front holography"~\cite{deTeramond:2008ht},  is  now providing a  color-confining approximation to $H_{LF}^{QCD}$ for QCD(3+1).  This method gives a remarkable first approximation to  hadron spectroscopy and  hadronic LFWFs.   A new method for solving nonperturbative QCD ``Basis Light-Front Quantization" (BLFQ)~\cite{Vary:2014tqa},  uses the eigensolutions of a color-confining approximation to QCD (such as LF holography) as the basis functions,  rather than the plane-wave basis used in DLCQ, thus incorporating the full dynamics of QCD.  LFWFs can also be determined from the covariant Bethe-Salpeter wavefunction by integrating over $k^-$~\cite{Brodsky:2015aia}.  
A review of the light-front formalism is given in Ref.~\cite{Brodsky:1997de}.

\section{Color Confinement and Supersymmetry in Hadron Physics from LF Holography}

A fundamental problem in hadron physics is to obtain a  color-confining first approximation to QCD which can predict both the hadron spectrum and the LFWFs underlying hadron phenomenology.  The QCD Lagrangian with zero quark mass has no explicit mass scale;   the classical theory is  conformally invariant.
A  profound question is  then to understand how the proton mass and other hadronic mass scales -- the mass gap -- can arise even when $m_q=0.$  In fact, chiral QCD has no knowledge of units such as $MeV$. 
However, a remarkable principle,  first demonstrated by  de Alfaro, Fubini and Furlan  (dAFF)~\cite{deAlfaro:1976je} in $1+1$ quantum mechanics, is that a mass scale can appear in a Hamiltonian without 
affecting  the conformal invariance of the action.  The essential step is to add to the conformal Hamiltonian $H_0$ terms proportional to the dilation operator $D$ and the special conformal operator $K$. The coefficients introduce the mass scale $\kappa$, and the result is  $H= H_0+ V$, where $V$ a  harmonic oscillator potential $V(x) = \kappa^2 x^2$.  The action remains conformal when one changes to a new time variable. 
De T\'eramond, Dosch, and I~\cite{Brodsky:2013ar}
have shown that a mass gap and a fundamental color confinement scale appear when one extends the dAFF procedure to light-front Hamiltonian theory.   
Remarkably, the resulting light-front potential has a unique form of a harmonic oscillator $\kappa^4 \zeta^2$ in the 
light-front invariant impact variable $\zeta$ where $ \zeta^2Ê = b^2_\perp x(1-x)$. The result is  a single-variable frame-independent relativistic equation of motion for  $q \bar q $ bound states, a ``Light-Front Schr\"odinger Equation"~\cite{deTeramond:2008ht}, analogous to the nonrelativistic radial Schr\"odinger equation in quantum mechanics. The same result, including spin terms, is obtained using  light-front holography  -- the duality between the front form and AdS$_5$, the space of isometries of the conformal group -- if one  
modifies the action of AdS$_5$ by the dilaton $e^{\kappa^2 z^2}$ in the fifth dimension $z$.  The  Light-Front Schr\"odinger Equation  incorporates color confinement and other essential spectroscopic and dynamical features of hadron physics, including a massless pion for zero quark mass and linear Regge trajectories with the same slope  in the radial quantum number $n$   and internal  orbital angular momentum $L$.      When one generalizes this procedure using superconformal algebra, the resulting light-front eigensolutions predict a unified Regge spectroscopy of meson, baryon, and tetraquarks, including remarkable supersymmetric relations between the masses of mesons and baryons of the same parity.

\begin{figure}
 \begin{center}
\includegraphics[height= 12cm,width=15cm]{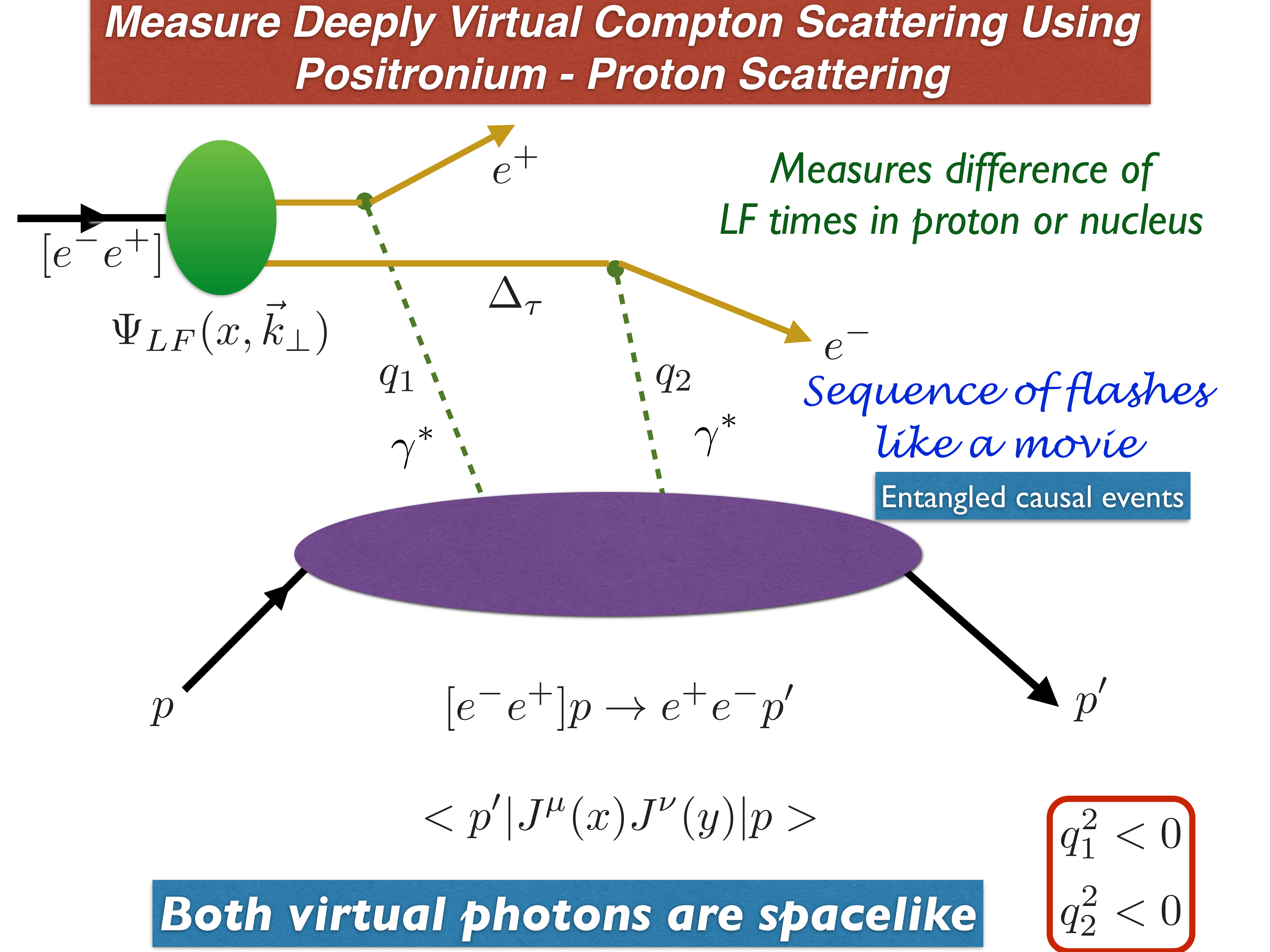}
\end{center}
\caption{Doubly Virtual Compton scattering on a proton (or nucleus) can be measured for two {\it spacelike} photons $q^2_1, q^2_2 <0$ 
with minimal, tunable, skewness $\xi$ using positronium-proton scattering $[e^+ e^-] p \to e^+ e^- p'$.  
One can also measure double deep inelastic scattering and elastic positronium-proton scattering.  
An analogous process will create the ``true muonium" atom $[\mu^- \mu^-]$~\cite{Brodsky:2009gx,Banburski:2012tk}.
 }
\label{Positronium}
\end{figure}

An essential point is that the mass scale $\kappa$ is not determined absolutely by QCD.    Only ratios of masses are determined,  and  the theory has dilation invariance under $\kappa \to C \kappa, $ In a sense, chiral QCD has an ``extended conformal invariance."  The resulting new time variable  which retains the conformal invariance of  the action, has finite support, conforming to the fact that the LF time between the interactions with the confined constituents is finite.  
The finite time difference $\Delta \tau$ between the LF times  $\tau$ of the quark constituents of the proton could be measured using positronium proton scattering $[e^+ e^-] p \to e^+ e^- p'$.  This process, which measures double diffractive deeply virtual Compton scattering for two spacelike photons, is illustrated in Fig.~\ref{Positronium}  One can also study the dissociation of relativistic positronium
atoms to an electron and positron with light front momentum fractions $x$ and $1-x$ and  opposite transverse momenta in analogy to the E791 measurements of the diffractive dissociation of the pion to two jets~\cite{Ashery:2000yj}
The LFWF of positronium in the relativistic domain is the central input.
One can produce a relativistic positronium beam  using the collisions of laser photons with high energy photons or by 
using Bethe-Heitler pair production below the $e^+ e^-$ threshold.
The production of parapositronium via the collision of photons is analogous to pion production in two-photon collisions and Higgs production via gluon-gluon fusion.

\section{Light-Front Holography} 
 
\begin{figure}
 \begin{center}
\includegraphics[height= 12cm,width=15cm]{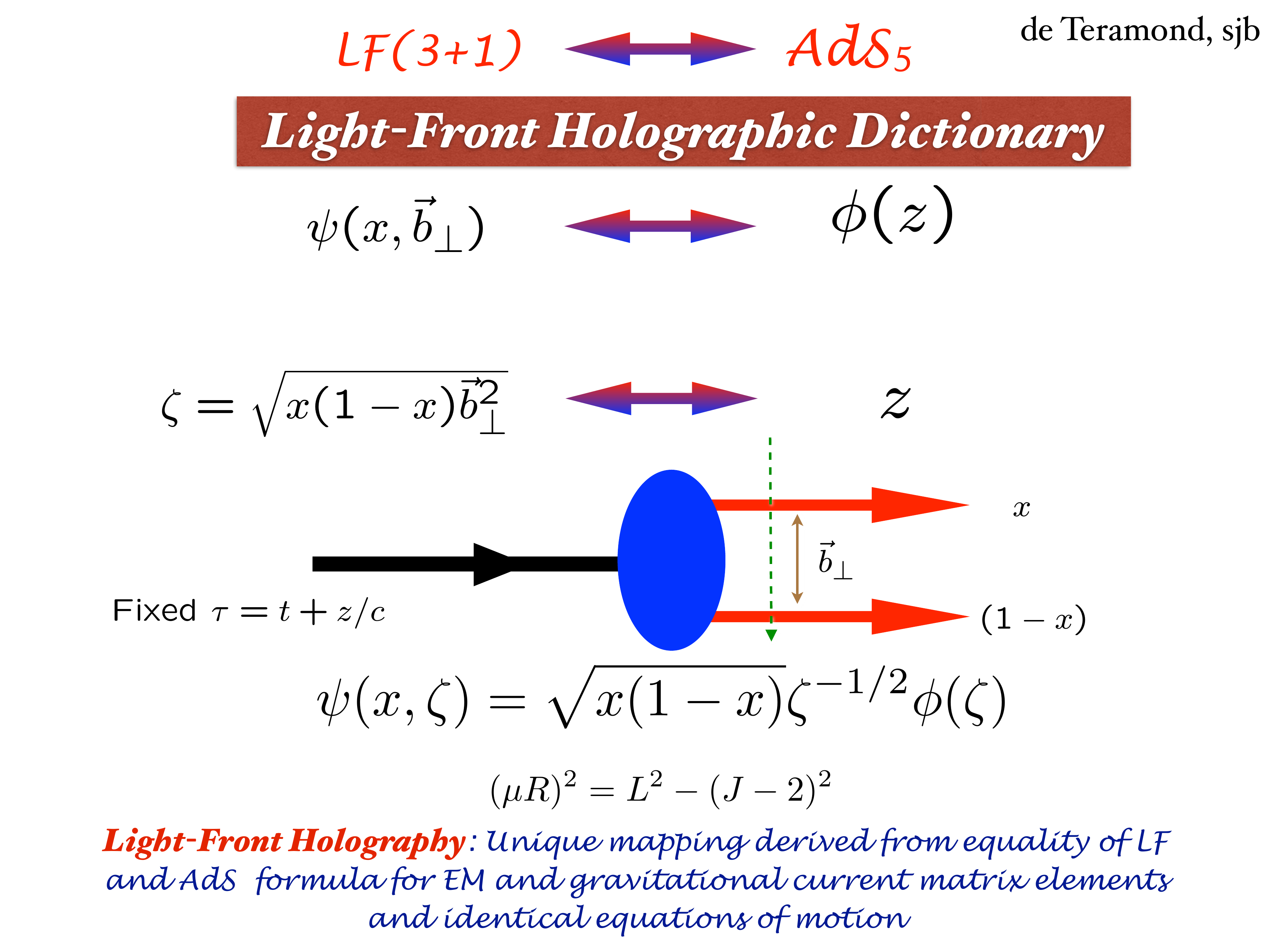}
\end{center}
\caption{The holographic dictionary which maps the fifth dimension variable $z$ of  the five-dimensional AdS$_5$ space to the LF radial variable $\zeta$ where 
$\zeta^2  =  b^2_\perp(1-x)$.   The same physics transformation maps the AdS$_5$  and $(3+1)$ LF expressions for electromagnetic and gravitational form factors to each other. 
From ref.~\cite{deTeramond:2013it}}
\label{dictionary}
\end{figure}

\begin{figure}
 \begin{center}
\includegraphics[height= 12cm,width=15cm]{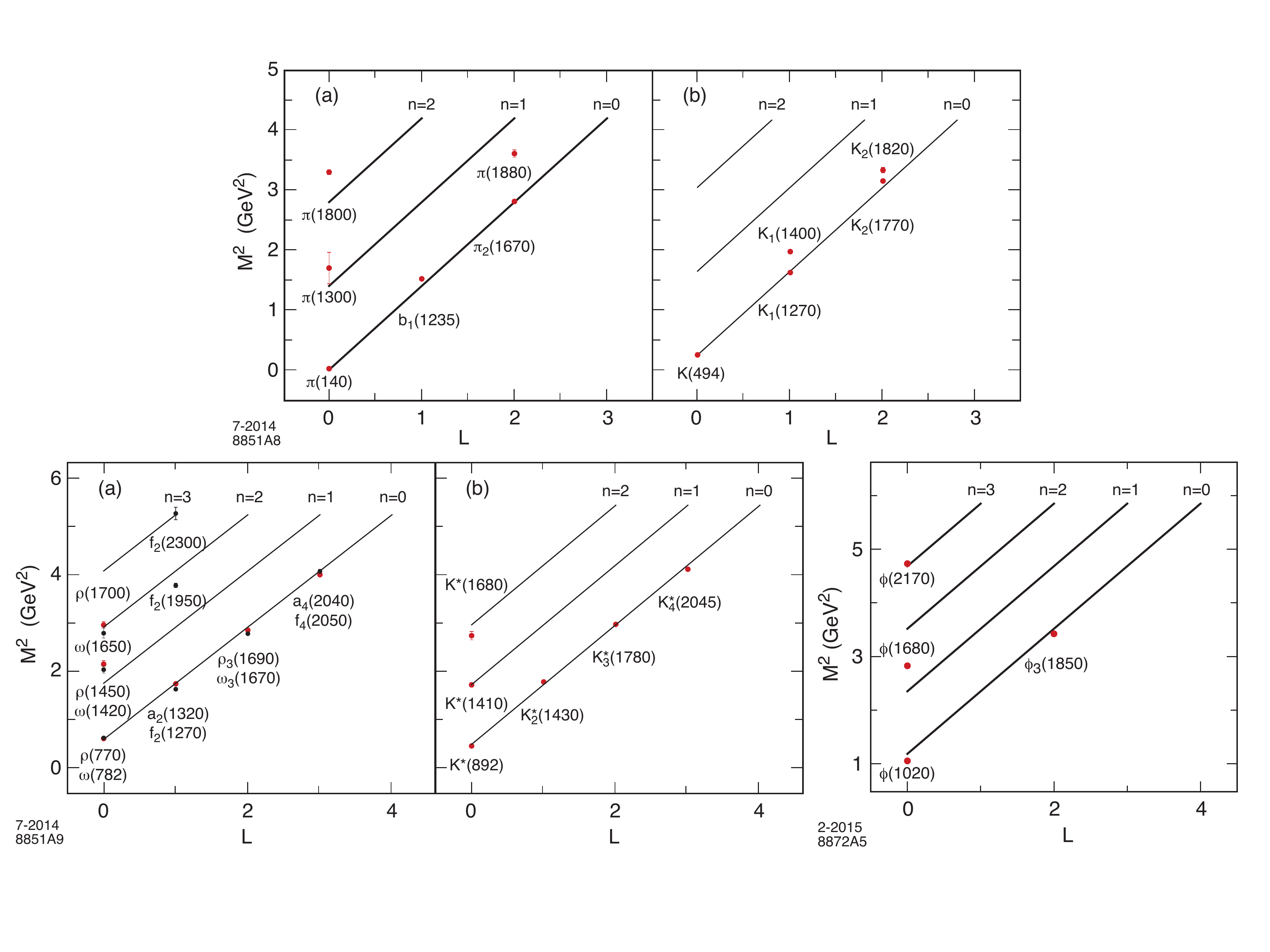}
\end{center}
\caption{Comparison of the AdS/QCD prediction  $M^2(n, L, S) = 4\kappa^2(n+L +S/2)$ for the orbital $L$ and radial $n$ excitations of the meson spectrum with experiment.   The pion is predicted to be massless for zero quark mass. The $u,d,s$ quark masses can be taken into account by perturbing in $<m_q^2/x>$.   The fitted value of $\kappa = 0.59$ MeV for pseudoscalar mesons, 
and  $\kappa = 0.54$ MeV  for vector mesons. }
\label{ReggePlot.pdf}
\end{figure}

Five-dimensional AdS$_5$ space provides a geometrical representation of the conformal group.
The identical color-confining light-front  equation for mesons of arbitrary spin $J$ can be derived~\cite{deTeramond:2013it}
from the holographic mapping of  the ``soft-wall model" modification of AdS$_5$ space for the specific dilaton profile $e^{+\kappa^2 z^2},$  where one identifies the fifth dimension coordinate $z$ with the light-front coordinate $\zeta$.  
Remarkably ,  AdS$_5$  is holographically dual to $3+1$  spacetime at fixed light-front time $\tau = t+ z/c$.  
The holographic dictionary is summarized in Fig. \ref{dictionary} 
The  combination of light-front dynamics, its holographic mapping to AdS$_5$ space, and the dAFF procedure provides new  insight into the physics underlying color confinement, the nonperturbative QCD coupling, and the QCD mass scale.  A comprehensive review is given in  Ref.~\cite{Brodsky:2014yha}.  The $q \bar q$ mesons and their valence LF wavefunctions are the eigensolutions of the frame-independent relativistic bound state LF Schr\"odinger equation.  
The mesonic $q\bar  q$ bound-state eigenvalues for massless quarks are $M^2(n, L, S) = 4\kappa^2(n+L +S/2)$.
The equation predicts that the pion eigenstate  $n=L=S=0$ is massless at zero quark mass. The  Regge spectra of the pseudoscalar $S=0$  and vector $S=1$  mesons  are 
predicted correctly, with equal slope in the principal quantum number $n$ and the internal orbital angular momentum $L$.  A comparison with experiment is shown in Fig. \ref{ReggePlot.pdf}.

\begin{figure}
\begin{center}
\includegraphics[height=10cm,width=15cm]{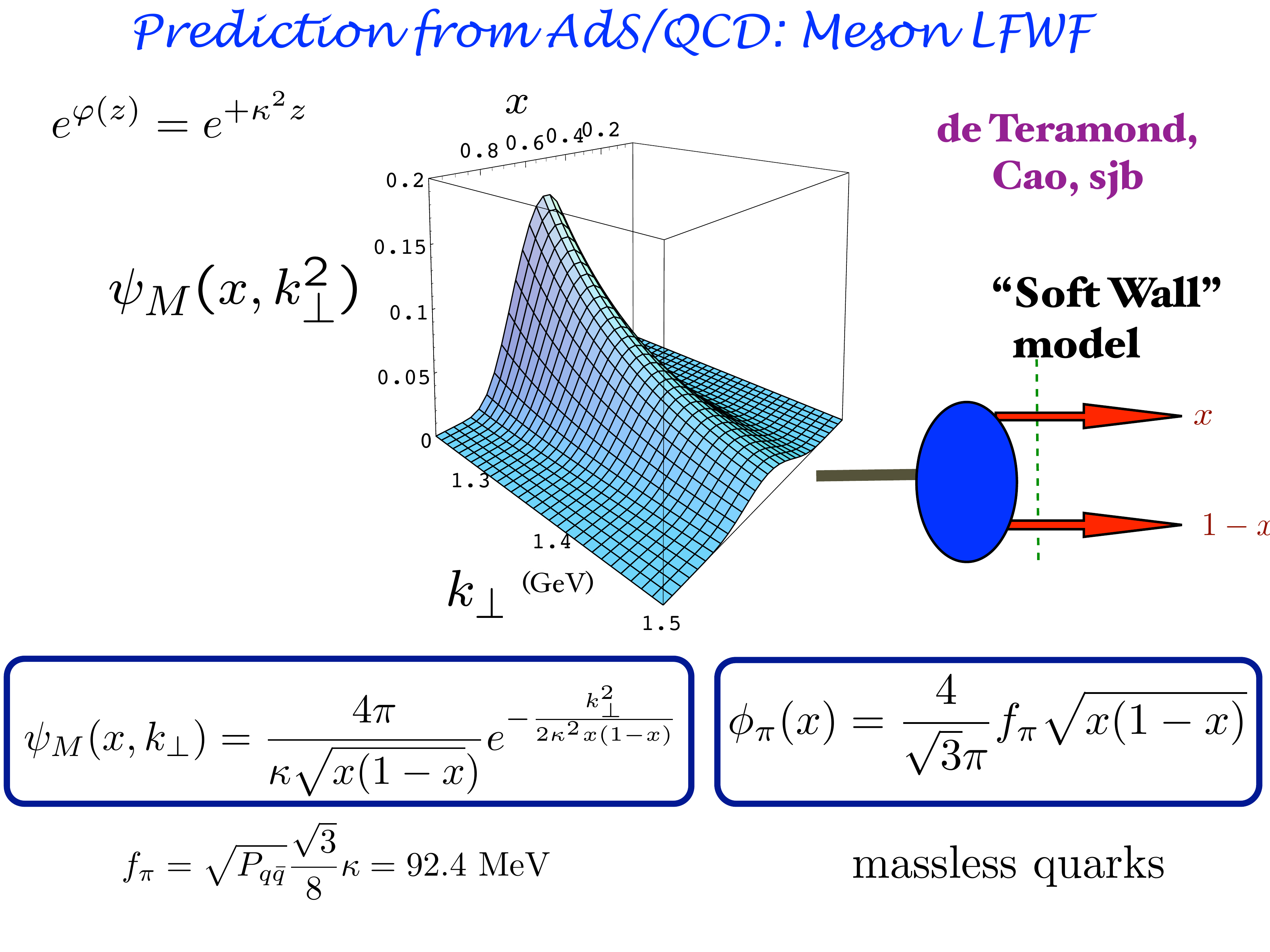}
\end{center}
\caption{Prediction from AdS/QCD and Light-Front Holography for  meson LFWFs  $\psi_M(x,\vec k_\perp)$   and the pion distribution amplitude.     
}
\label{MesonLFWF.pdf}
\end{figure}

The predicted hadronic LFWFs are  functions of the LF kinetic energy $\vec k^2_\perp/ x(1-x)$ -- the conjugate of the LF radial variable $\zeta^2 = b^2_\perp x(1-x)$ -- times a function of $x(1-x)$; they do not factorize as a  function of $\vec k^2_\perp$ times a function of $x$.  The resulting  nonperturbative pion distribution amplitude $\phi_\pi(x) = \int d^2 \vec k_\perp \psi_\pi(x,\vec k_\perp) = (4/  \sqrt 3 \pi)  f_\pi \sqrt{x(1-x)}$,  see Fig. \ref{MesonLFWF.pdf}, which controls hard exclusive process, is  consistent with the Belle data for the photon-to-pion transition form factor~\cite{Brodsky:2011xx}.  
The AdS/QCD light-front holographic eigenfunction for the $\rho$ meson LFWF $\psi_\rho(x,\vec k_\perp)$ gives excellent 
predictions for the observed features of diffractive $\rho$ electroproduction $\gamma^* p \to \rho  p^\prime$,  as shown by Forshaw and Sandapen~\cite{Forshaw:2012im}

It is interesting to note that the contribution of the {\it `H'} diagram to $Q \bar Q $ scattering is IR divergent as the transverse separation between the $Q$  
and the $\bar Q$ increases~\cite{Smirnov:2009fh}.  This is a signal that pQCD is inconsistent without color confinement.  The sum of such diagrams could sum to the confinement potential $\kappa^4 \zeta^2 $ dictated by the dAFF principle that the action remains conformally invariant despite the appearance of the mass scale $\kappa$ in the Hamiltonian.
The $\kappa^4 \zeta^2$ confinement interaction between a $q$ and $\bar q$ will induce a $\kappa^4/s^2$ correction to $R_{e^+ e^-}$, replacing the $1/ s^2$ signal usually attributed to a vacuum gluon condensate.  

.

\section{Supersymmetric  Hadron Physics}

\begin{figure}
 \begin{center}
\includegraphics[height=10cm,width=15cm]{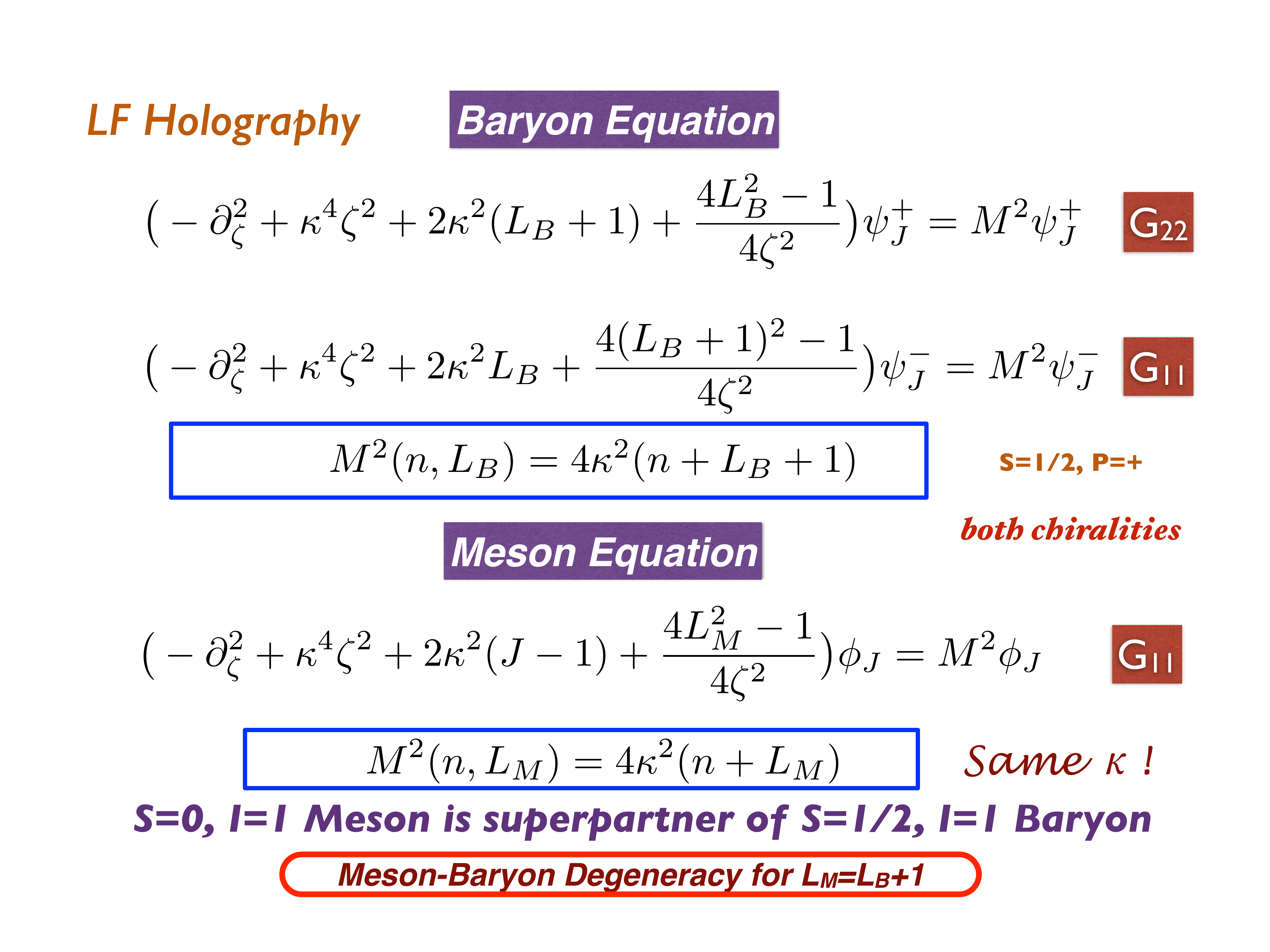}
\includegraphics[height=10cm,width=15cm]{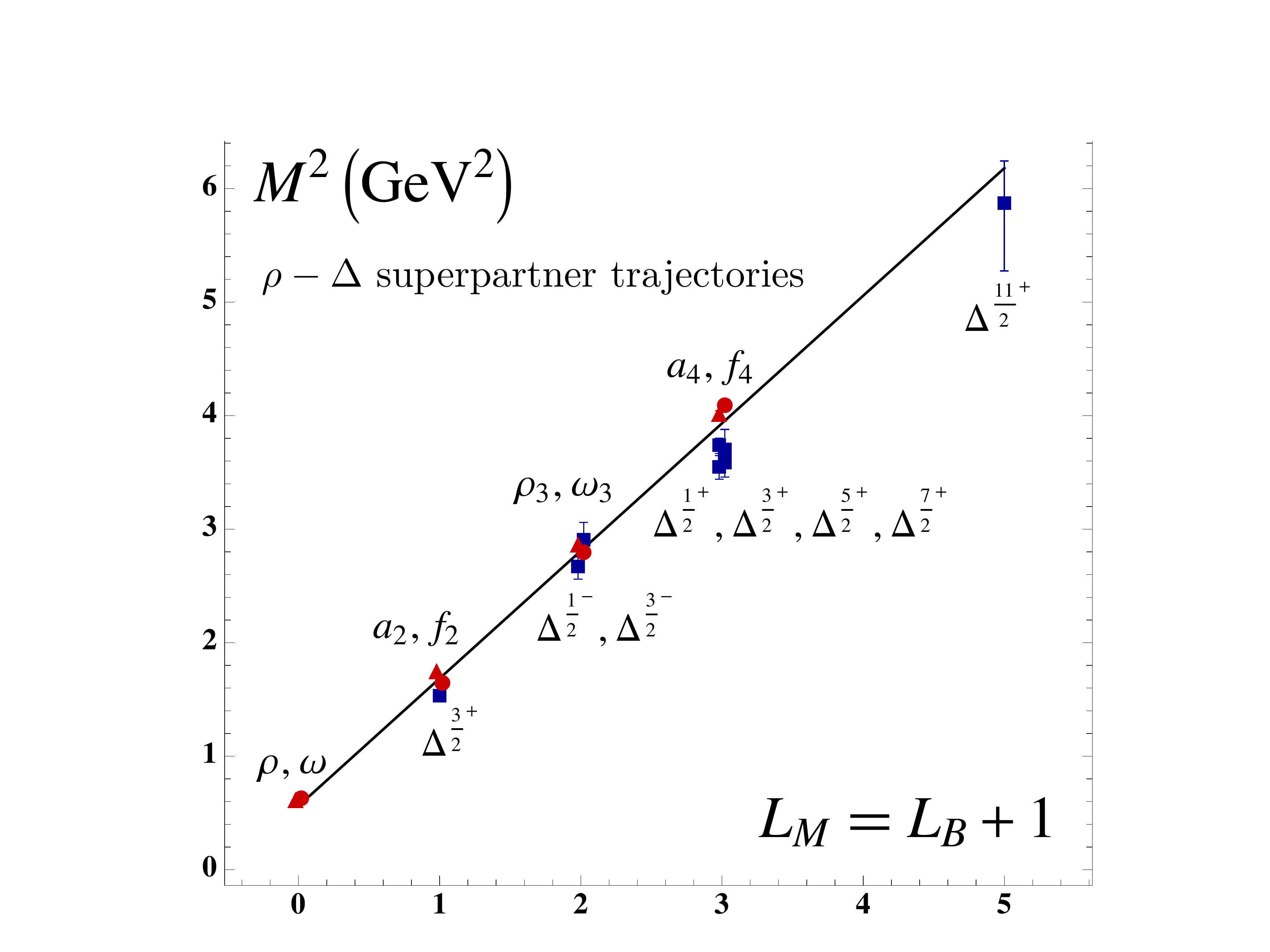}
\end{center}
\caption{(A). The LF Schr\"odinger equations for baryons and mesons for zero quark mass derived from the Pauli $2\times 2$ matrix representation of superconformal algebra.  
The $\psi^\pm$  are the baryon quark-diquark LFWFs where the quark spin $S^z_q=\pm 1/2$ is parallel or antiparallel to the baryon spin $J^z=\pm 1/2$.   The meson and baryon equations are identical if one identifies a meson with internal orbital angular momentum $L_M$ with its superpartner baryon with $L_B = L_M-1.$
See Refs.~\cite{deTeramond:2014asa,Dosch:2015nwa,Dosch:2015bca}.
(B). Comparison of the $\rho/\omega$ meson Regge trajectory with the $J=3/2$ $\Delta$  baryon trajectory.   Superconformal algebra  predicts the degeneracy of the  meson and baryon trajectories if one identifies a meson with internal orbital angular momentum $L_M$ with its superpartner baryon with $L_M = L_B+1.$
See Refs.~\cite{deTeramond:2014asa,Dosch:2015nwa}.}
\label{FigsJlabProcFig3.pdf}
\end{figure}

The conformal group has an elegant $ 2\times 2$ Pauli matrix representation called superconformal algebra, originally discovered by  Haag, Lopuszanski, and Sohnius ~\cite{Haag:1974qh}(1974)
The conformal Hamiltonian operator and the special conformal operators can be represented as anticommutators of Pauli matrices
 $H = {1/2}[Q, Q^\dagger]$ and  $K = {1/2}[S, S^\dagger]$.
As shown by Fubini and Rabinovici,~\cite{Fubini:1984hf},  a nonconformal Hamiltonian with a mass scale and universal confinement can then be obtained by shifting $Q \to Q +\omega K$, the analog of the dAFF procedure. 
In effect one has generalized supercharges of the superconformal algebra~\cite{Fubini:1984hf}. 
The resulting superconformal algebra leads to effective QCD light-front  bound-state equations for both mesons and baryons~\cite{deTeramond:2014asa,Dosch:2015nwa,Dosch:2015bca}.  
The supercharges connect the baryon and meson spectra  and their Regge trajectories to each other in a remarkable manner: each meson has internal  angular momentum one unit higher than its superpartner baryon  $L_M = L_B+1.$  The resulting set of LF equations for confined quarks are shown in Fig. \ref {FigsJlabProcFig3.pdf}(A). 
In effect the baryons are color-singlet bound-states of color-triplet quarks and $\bar 3_C$ $[qq]$ diquarks which are themselves $3_C \times 3_C$ clusters.
Note that the same slope controls the Regge trajectories of both mesons and baryons in both the orbital angular momentum $L$ and the principal quantum number $n$.
Only one mass parameter $\kappa = \omega^2$  appears; it sets the confinement and the hadron mass scale in the  chiral limit, as well as  the length scale which underlies hadron structure.  ``Light-Front Holography"  not only predicts meson and baryon  spectroscopy  successfully, but also hadron dynamics, including  vector meson electroproduction,  hadronic light-front wavefunctions, distribution amplitudes, form factors, and valence structure functions.  
The LF Schr\"odinger Equations for baryons and mesons derived from superconformal algebra  are shown  in Fig. \ref{FigsJlabProcFig3.pdf}.
In effect the baryons on the proton (Delta) trajectory are bound states of a quark with color $3_C$ and scalar (vector)  diquark with color $\bar 3_C$ 
The proton eigenstate labeled $\psi^+$ (parallel quark and baryon spins) and $\psi^-$ (anti parallel quark and baryon spins)  have equal Fock state probability -- a  feature of ``quark chirality invariance".  Predictions for the static properties of the nucleons are discussed in ref. ~\cite{Liu:2015jna}

The comparison between the meson and baryon masses of the $\rho/\omega$ Regge trajectory with the spin-$3/2$ $\Delta$ trajectory 
is shown in Fig. \ref{FigsJlabProcFig3.pdf}(B).
Superconformal algebra  predicts that the bosonic meson and fermionic baryon masses are equal if one identifies each meson with internal orbital angular momentum $L_M$ with its superpartner baryon with $L_B = L_M-1$; the meson and baryon  superpartners  then have the same parity,  Since
 $ 2+ L_M = 3 + L_B$, the meson and baryon superpartners are also the same.   Superconformal algebra also predicts that the LFWFs of the superpartners are identical, and thus the corresponding elastic and transition form factors are equal.   The predicted identity of meson and baryon timelike form factors can be tested in $e^+ e^- \to H \bar H^\prime $ reactions. 

\begin{figure}
\begin{center}
\includegraphics[height=10cm,width=15cm]{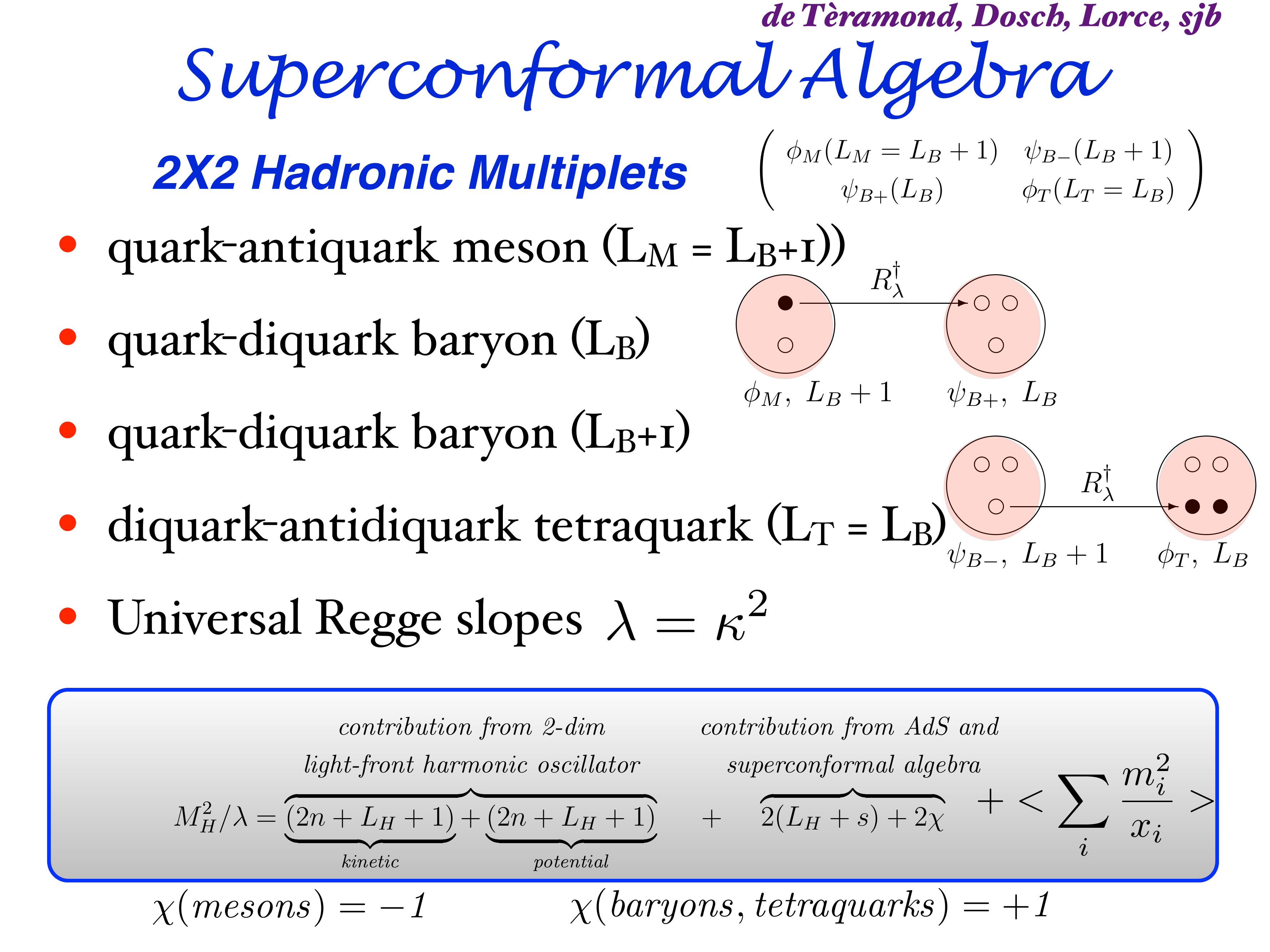}
\end{center}
\caption{The eigenstates of superconformal algebra have a $2 \times 2 $ representation of mass degenerate bosons and fermions:  a meson with $L_M=  L_B+1$, a baryon doublet with 
$L_B, L_B+1$ components and a tetraquark with $ L_T = L_B$. The breakdown of LF kinetic, potential, spin, and quark mass contributions to each hadron is also shown.  The virial theorem predicts the equality of the LF kinetic and potential contributions.
}
\label{2X2Multiplets}
\end{figure}

As illustrated in fFg. \ref{2X2Multiplets}, the hadronic eigensolutions of the superconformal algebra are  themselves  $2\times 2$ matrices connected internally by the supersymmetric algebra operators. In addition to the meson and baryon eigenstates, one also predicts color singlet {\it tetraquark}  diquark-antidiquark bound states with the same mass as the baryon.
One can also generalize these results to heavy-light $[qQ] $ mesons and  $[[qq]Q]$ baryons.  The mass parameter controlling the Regge slopes will be increased for heavy $m_Q$; however, the supersymmetric connections between the heavy-light hadrons is predicted to be maintained.

\section{Ridge formation from flux-tube collisions}

In the case of $e p \to e^\prime X$, one can consider the collisions of the confining  QCD flux tube appearing between the $q$ and $\bar q$  of the virtual photon with the flux tube between the quark and diquark of the proton.   Since the $q \bar q $ plane is aligned with the scattered electron's plane, the resulting ``ridge"  of hadronic multiplicity produced from the $\gamma^* p$ collision will also tend to be aligned with the scattering plane of the scattered electron.  The virtual photon's flux tube will also depend on the photon virtuality $Q^2$, as well as the flavor of the produced pair arising from $\gamma^* \to q \bar q$.  In the case of high energy $\gamma^* \gamma^*$ collisions, one can control the  produced  hadron multiplicity and ridge geometry using the scattered electrons' planes or the scattered proton planes in ultra-peripheral collisions at the  LHC. The resulting dynamics~\cite{Brodsky:2014hia} is  a natural extension of the flux-tube collision description of the ridges produced in $p-p$ collisions~\cite{Bjorken:2013boa}.

\section{Calculations using LF-Time-Ordered Perturbation Theory and Hadronization at the Amplitude Level}

LF-time-ordered perturbation theory  can be advantageous  for perturbative QCD calculations.  An excellent example  of LF-time-ordered perturbation theory is the computation of multi-gluon scattering amplitudes by Cruz-Santiago and Stasto~\cite{Cruz-Santiago:2015dla}.  
  In this method, the propagating particles  are on their respective mass shells:  $k_\mu k^\mu = m^2$, and intermediate states are off-shell in invariant mass;  {\it i.e.}, $P^- \ne \sum k^-_i$. Unlike instant form, where one must sum  $n !$ frame-dependent  amplitudes, only  the $\tau$-ordered diagrams where each propagating particle has  positive $k^+ =k^0+k^z$  can contribute. 
The number of nonzero amplitudes is also greatly reduced by noting that the total angular momentum projection $J^z = \sum_i^{n-1 } L^z_i + \sum^n_i S^z_i$ and the total $P^+$ are  conserved at each vertex.  In a renormalizable theory, the change in orbital angular momentum is limited to $\Delta L^z =0,\pm 1$ at each vertex~\cite{Kelly}

A remarkable advantage of LF time-ordered perturbation theory (LFPth) is that the calculation of a subgraph of any order in pQCD only needs to be done once;  the result can be stored in a ``history" file.  This is due to the fact that in LFPth the numerator algebra is independent of the process; the denominator changes, but only by a simple shift of the initial $P^-$.   Another simplification is that loop integrations are three dimensional: $\int d^2\vec k_\perp \int^1_0 dx.$   Unitarity  and explicit
 renormalization can be implemented using the ``alternate denominator" method which defines the required subtraction counterterms~\cite{Brodsky:1973kb}.
 
The new insights into color confinement given by AdS/QCD suggest that one could compute hadronization at  amplitude level~\cite{Brodsky:2009dr} using  the confinement interaction and the LFWFs predicted by  AdS/QCD and Light-Front Holography.  One can postulate that  the invariant mass of a color-singlet cluster ${\cal M}$  is the variable which separates perturbative and nonperturbative dynamics.
For example, consider $e^+ e^- $ annihilation using LF $\tau$ - ordered perturbation theory.   
At an early stage in LF time  the annihilation will produce jets of quarks and gluons in an intermediate state that are off the $P^-$ energy shell. 
If a color-singlet cluster of partons in a jet satisfies ${\cal M}^2 <  \kappa^2$,  
the cluster constituents will be ruled by the $\kappa^4\zeta^2$ color-confinement potential.
At this stage, the LFWF $\psi_H$ converts the off-shell partons to 
the on-shell hadron. Quarks and gluons only appear in intermediate states, but only hadrons can be produced.    
Thus the AdS/QCD Light-Front Holographic model  suggests how one can implement the transition between perturbative and nonperturbative QCD.  For a QED analog, see  Refs.~\cite{Brodsky:2009gx,Banburski:2012tk}.

\section{The Light-Front Vacuum}

It is important to distinguish the LF vacuum from the conventional instant-form vacuum.
The eigenstates of the instant-form Hamiltonian describe a state defined at a single instant of time $t$ over all space, and they are thus acausal as well as frame-dependent.  
The instant-form vacuum is defined as the lowest energy eigenstate of the instant-form Hamiltonian.
As discussed by Zee ~\cite{Zee:2008zz}, the cosmological constant  is of order $10^{120}$ times larger than what is observed if one computes the effects of quantum loops from QED.  Similarly, QCD instantons and condensates in the instant-form vacuum give a contribution of order  $10^{42}.$  The contribution of the  Higgs VEV computed in the instant form vacuum is  $10^{54}$ times too large. 

In contrast, the vacuum in LF Hamlitonian theory is defined as the eigenstate of $H_{LF}$ with lowest invariant mass.  It is defined at fixed LF time $\tau$ within the causal horizon, and  it is frame-independent; i.e., it is independent of the observer's motion.
Vacuum loop diagrams from quantum field theory do not appear  in the front-form vacuum since  the  $+$ momenta are positive: $k^+ _i = k^0_i+k^z_i\ge 0$, and the sum of $+$ momenta is conserved at every vertex.   The creation of particles cannot arise from the LF vacuum  since $ \sum_i  k^{+i}   \ne P^+_{vacuum} =0.$  
Since propagation with negative $k^+$  does not appear, the LF vacuum is trivial up to possible $k^+=0$ ``zero"  modes. The physics associated with quark and gluon QCD vacuum condensates of the instant form are replaced by physical effects contained within the hadronic LFWFs  in the hadronic domain. 
This is referred to as ``in-hadron" condensates~\cite{Casher:1974xd,Brodsky:2009zd,Brodsky:2010xf}.  
In the case of the Higgs theory, the traditional Higgs vacuum expectation value (VEV) is replaced by a ``zero mode", analogous to a classical 
Stark or Zeeman field~\cite{Srivastava:2002mw}.  
The Higgs LF zero mode~\cite{Srivastava:2002mw}  has no energy-momentum density,  so it also gives zero contribution to the cosmological constant.  

The universe is observed within the causal horizon, not at a single instant of time.  The causal, frame-independent light-front vacuum can thus provide a viable match to the empty visible universe~\cite{Brodsky:2010xf}.    The huge contributions to the cosmological constant  thus do not appear if one notes that the causal, frame-independent light-front vacuum has no quantum fluctuations --  in dramatic contrast to to  the acausal, frame-dependent instant-form vacuum;  the cosmological constant  arising from quantum field theory thus vanishes if one uses the front form.  The observed nonzero value could could be a property of gravity itself, such as the ``emergent gravity"  postulated by E. Verlinde~\cite{Verlinde:2016toy}. It is also possible that if one solves electroweak theory in a curved universe, the Higgs LF zero mode would be replaced with a field of nonzero curvature which could give a nonzero contribution.

\section {The QCD Coupling at all Scales} 

The QCD running coupling $\alpha_s(Q^2)$
sets the strength of  the interactions of quarks and gluons as a function of the momentum transfer $Q$.
The dependence of the coupling
$Q^2$ is needed to describe hadronic interactions at 
both long and short distances. 
The QCD running coupling can be defined~\cite{Grunberg:1980ja} at all momentum scales from a perturbatively calculable observable, such as the coupling $\alpha^s_{g_1}(Q^2)$, which is defined from measurements of the Bjorken sum rule.   At high momentum transfer, such ``effective charges"  satisfy asymptotic freedom, obey the usual pQCD renormalization group equations, and can be related to each other without scale ambiguity by commensurate scale relations~\cite{Brodsky:1994eh}.  

\begin{figure}
\begin{center}
\includegraphics[height=11cm,width=15cm]{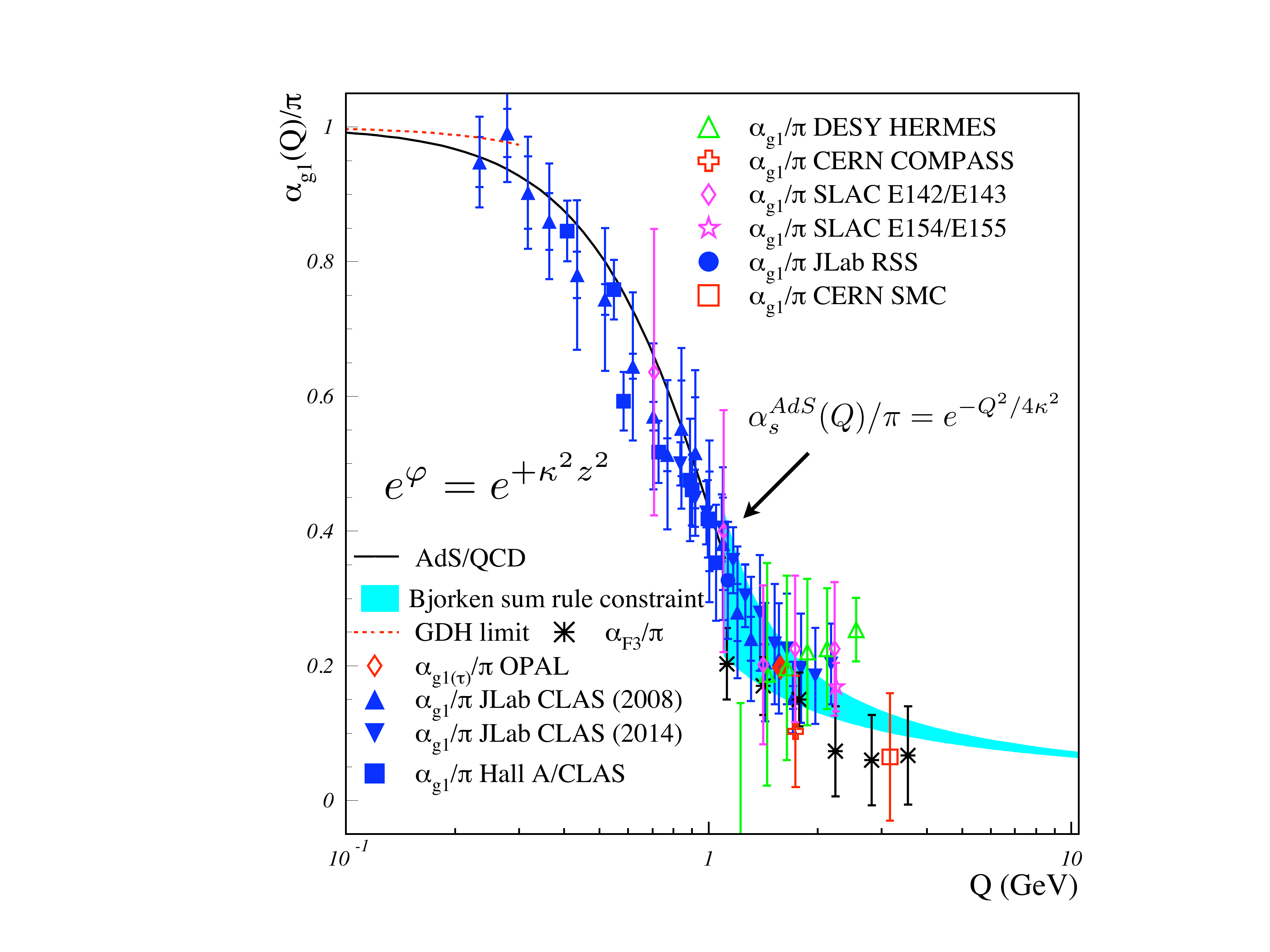}
\includegraphics[height=12cm,width=15cm]{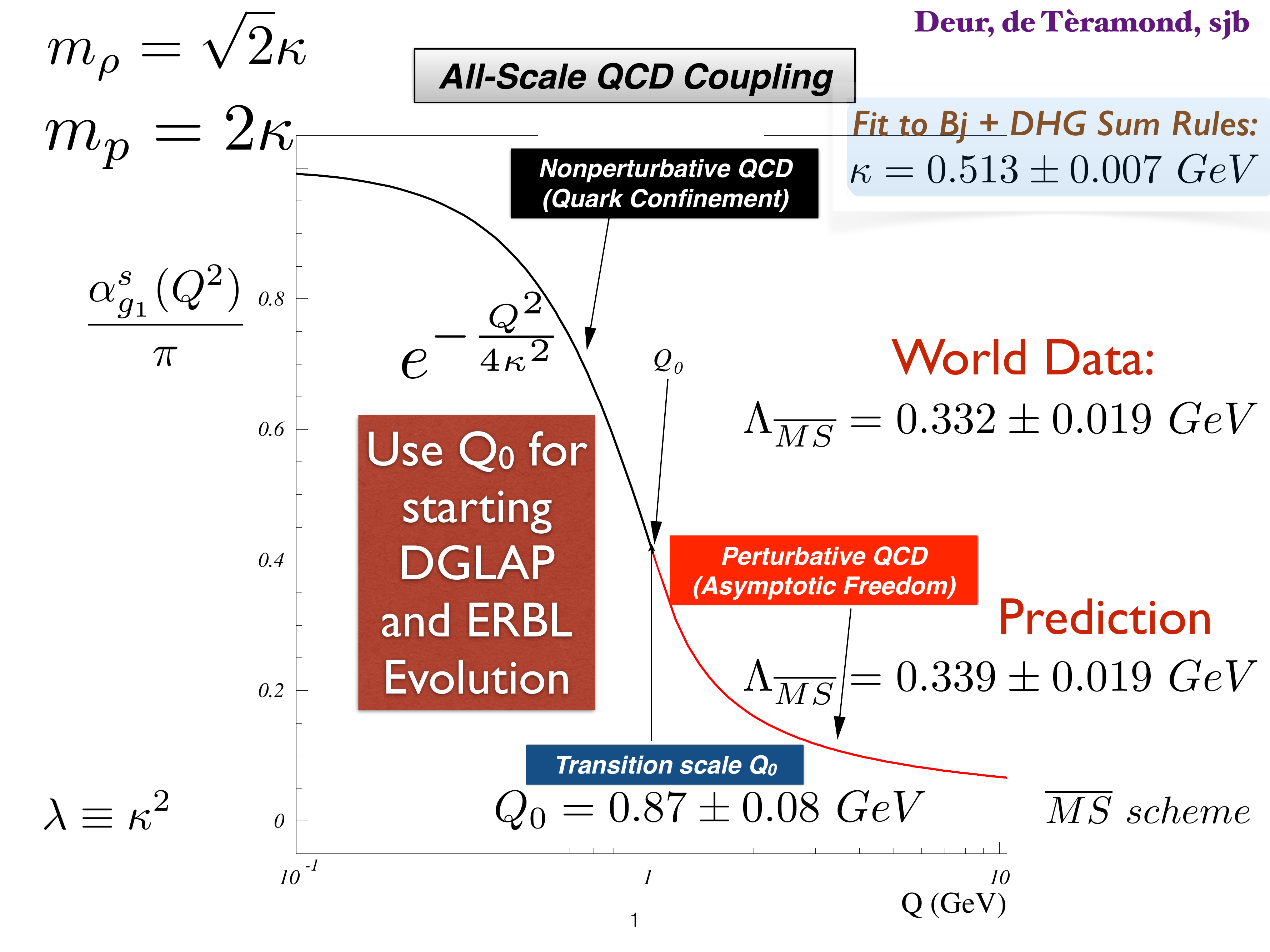}
\end{center}
\caption{
(A) Comparison of the predicted nonpertubative coupling, based on  the dilaton $\exp{(+\kappa^2 z^2)}$ modification of the AdS$_5$ metric, with measurements of the effective charge $\alpha^s_{g_1}(Q^2)$, 
as defined from the Bjorken sum rule.
(B)  Prediction from LF Holography and pQCD for the QCD running coupling $\alpha^s_{g_1}(Q^2)$ at all scales.   The magnitude and derivative of the perturbative and nonperturbative coupling are matched at the scale $Q_0$.  This matching connects the perturbative scale 
$\Lambda_{\overline{MS}}$ to the nonpertubative scale $\kappa$ which underlies the hadron mass scale. 
See Ref.~\cite{Brodsky:2014jia}. 
}
\label{DeurCoupling}
\end{figure}

The dilaton  $e^{+\kappa^2 z^2}$ soft-wall modification of the AdS$_5$ metric, together with LF holography, predicts the functional behavior of the running coupling
in the small $Q^2$ domain~\cite{Brodsky:2010ur}: 
${\alpha^s_{g_1}(Q^2) = 
\pi   e^{- Q^2 /4 \kappa^2 }}. $ 
Measurements of  $\alpha^s_{g_1}(Q^2)$ are remarkably consistent~\cite{Deur:2005cf}  with this predicted Gaussian form; the best fit gives $\kappa= 0.513 \pm 0.007~GeV$.   
See Fig.~\ref{DeurCoupling}
Deur, de Teramond, and I~\cite{Brodsky:2010ur,Deur:2014qfa,Brodsky:2014jia} have also shown how the parameter $\kappa$,  which   determines the mass scale of  hadrons and Regge slopes  in the zero quark mass limit, can be connected to the  mass scale $\Lambda_s$  controlling the evolution of the perturbative QCD coupling.  The high momentum transfer dependence  of the coupling $\alpha_{g1}(Q^2)$ is  predicted  by  pQCD.  The 
matching of the high and low momentum transfer regimes  of $\alpha_{g1}(Q^2)$ -- both its value and its slope -- then determines a scale $Q_0 =0.87 \pm 0.08$ GeV which sets the interface between perturbative and nonperturbative hadron dynamics.  This connection can be done for any choice of renormalization scheme, such as the $\overline{MS}$ scheme,
as seen in  Fig.~\ref{DeurCoupling}.  
The result of this perturbative/nonperturbative matching is an effective QCD coupling  defined at all momenta.   
The predicted value of $\Lambda_{\overline{MS}} = 0.339 \pm 0.019~GeV$ from this analysis agrees well the measured value~\cite{Agashe:2014kda}  
$\Lambda_{\overline{MS}} = 0.332 \pm 0.019~GeV.$
These results, combined with the AdS/QCD superconformal predictions for hadron spectroscopy, allow us to compute hadron masses in terms of $\Lambda_{\overline{MS}}$:
$m_p =  \sqrt 2 \kappa = 3.21~ \Lambda_{\overline{MS}},~ m_\rho = \kappa = 2.2 ~ \Lambda_{\overline{ MS} }, $ and $m_p = \sqrt 2 m_\rho, $ meeting a challenge proposed by Zee~\cite{Zee:2003mt}.
The value of $Q_0$ can be used to set the factorization scale for DGLAP evolution of hadronic structure functions and the ERBL evolution of distribution amplitudes.
Deur, de T\'eramond, and I have also computed the dependence of $Q_0$ on the choice of the  effective charge used to define the running coupling and the renormalization scheme used to compute its behavior in the perturbative regime.   
The use of  the scale $Q_0$  to  resolve  the factorization scale uncertainty in structure functions and fragmentation functions,  in combination with the scheme-indepedent {\it principle of maximum sensitivity} (PMC )~\cite{Mojaza:2012mf} for  setting   renormalization scales,  can 
greatly improve the precision of pQCD predictions for collider phenomenology.

\section{Is the Momentum Sum Rule Valid for Nuclear Structure Functions? }

Sum rules for deep inelastic scattering are usually analyzed using the operator product expansion of the forward virtual Compton amplitude, assuming it depends in the limit $Q^2 \to \infty$ on matrix elements of local operators such as the energy-momentum tensor.  The moments of structure functions and other distributions can then be evaluated as overlaps of the target hadron's light-front wavefunction,  as in the Drell-Yan-West formulae for hadronic form factors~\cite{Brodsky:1980zm,Liuti:2013cna,Mondal:2015uha,Lorce:2011dv}.
The real phase of the resulting DIS amplitude and its OPE matrix elements reflects the real phase of the stable target hadron's wavefunction.

The ``handbag" approximation to deeply virtual Compton scattering also defines the ``static"  contribution~\cite{Brodsky:2008xe,Brodsky:2009dv} to the measured parton distribution functions (PDF), transverse momentum distributions, etc.  The resulting momentum, spin and other sum rules reflect the properties of the hadron's light-front wavefunction.
However, final-state interactions which occur {\it after}  the lepton scatters on the quark, can give non-trivial contributions to deep inelastic scattering processes at leading twist and thus survive at high $Q^2$ and high $W^2 = (q+p)^2.$  For example, the pseudo-$T$-odd Sivers effect~\cite{Brodsky:2002cx} is directly sensitive to the rescattering of the struck quark. 
Similarly, diffractive deep inelastic scattering (DDIS)  involves the exchange of a gluon after the quark has been struck by the lepton~\cite{Brodsky:2002ue}.  In each case the corresponding DVCS amplitude is not given by the handbag diagram since interactions between the two currents are essential.
These ``lensing" corrections survive when both $W^2$ and $Q^2$ are large since the vector gluon couplings grow with energy.  Part of the final state phase can be associated with a Wilson line as an augmented LFWF~\cite{Brodsky:2010vs} which does not affect the moments.  

The Glauber propagation  of the vector system $V$ produced by the  DDIS interaction on the nuclear front face and its subsequent  inelastic interaction with the nucleons in the nuclear interior $V + N_b \to X$ occurs after the lepton interacts with the struck quark.  The corresponding DVCS amplitude is not given by the handbag diagram since interactions between the two currents are essential.
Because of the rescattering dynamics, the DDIS amplitude acquires a complex phase from Pomeron and Regge exchange;  thus final-state  rescattering corrections lead to  nontrivial ``dynamical" contributions to the measured PDFs; i.e., they are a consequence of the scattering process itself~\cite{Brodsky:2013oya}.  The $ I = 1$ Reggeon contribution to DDIS on the front-face nucleon then leads to flavor-dependent antishadowing~\cite{Brodsky:1989qz,Brodsky:2004qa}.  This could explain why the NuTeV charged current measurement $\mu A \to \nu X$ scattering does not appear to show antishadowing, in contrast to deep inelastic electron-nucleus scattering as discussed in ref. ~\cite{Schienbein:2007fs}.

Diffractive deep inelastic scattering is leading-twist. and it is an essential component of the two-step amplitude which causes shadowing and antishadowing of the nuclear PDF.  It is important to analyze whether the momentum and other sum rules derived from the OPE expansion in terms of local operators remain valid when these dynamical rescattering corrections to the nuclear PDF are included.   The OPE is derived assuming that the LF time separation between the virtual photons in the forward virtual Compton amplitude 
$\gamma^* A \to \gamma^* A$  scales as $1/Q^2$.
However, the propagation  of the vector system $V$ produced by the DDIS interaction on the front face and its inelastic interaction with the nucleons in the nuclear interior $V + N_b \to X$ are characterized by a non-vanishing LF time  interval in the nuclear rest frame.   
Note also  that shadowing in deep inelastic lepton scattering on a nucleus  involves  nucleons facing the incoming lepton beam. The  geometrical orientation of the shadowed nucleons  is not a property the frame-independent nuclear LFWFs used to evaluate the matrix elements of local currents.  Thus leading-twist shadowing and antishadowing appear to invalidate the sum rules for nuclear PDFs.  The same complications occur in the leading-twist analysis of deeply virtual Compton scattering $\gamma^* A \to \gamma^* A$ on a nuclear target. Thus the leading-twist multi-nucleon processes which produce shadowing and antishadowing in a nucleus are  not accounted for using the $Q^2 \to \infty$ OPE analysis.

\section{Summary} Light-Front Quantization provides a physical, frame-independent formalism for hadron dynamics and structure.  Observables such as structure functions, transverse momentum distributions, and distribution amplitudes are defined from the hadronic light-front wavefunctions.   One obtains new insights into the hadronic 
spectrum, light-front wavefunctions, and the $e^{-{Q^2\over 4 \kappa^2}}$ Gaussian functional form of the QCD running coupling in the nonperturbative domain using light-front holography -- the duality between the front form   and AdS$_5$, the space of isometries of the conformal group. 
In addition, superconformal algebra leads to remarkable supersymmetric relations between mesons and baryons of the same parity.  The mass scale $\kappa$ underlying confinement and hadron masses  can be connected to the parameter   $\Lambda_{\overline {MS}}$ in the QCD running coupling by matching the nonperturbative dynamics, as described by  the effective conformal theory mapped to the light-front and its embedding in AdS space, to the perturbative QCD  regime. The result is an effective coupling  defined at all momenta.   This 
matching of the high and low momentum transfer regimes determines a scale $Q_0$ which  sets the interface between perturbative and nonperturbative hadron dynamics.  
The use of $Q_0$ to  resolve  the factorization scale uncertainty for structure functions and distribution amplitudes,  in combination with the principle of maximal conformality (PMC)  for  setting the  renormalization scales,  can 
greatly improve the precision of perturbative QCD predictions for collider phenomenology.  The absence of vacuum excitations of the causal, frame-independent  front form vacuum has important consequences  for the cosmological constant.   I have also discussed evidence that the antishadowing of nuclear structure functions is non-universal; {\it i.e.},  flavor dependent, and why shadowing and antishadowing phenomena may be incompatible with sum rules for nuclear parton distribution functions.

\section*{Acknowledgments}

Presented at Light-Cone 2016,{\it Challenges  in Experimental and Theoretical Physics on the Light Front} September 5-8, 2015,  IST, Universidade de Lisboa,  Lisbon, Portugal.
The results presented here are based on collaborations  and discussions  with  
James Bjorken, Kelly Chiu, Alexandre Deur, Guy de T\'eramond, Guenter Dosch, Susan Gardner, Fred Goldhaber,  Paul Hoyer, Dae Sung Hwang,  Rich Lebed, 
Simonetta Liuti, Cedric Lorce, Matin Mojaza,  Michael Peskin, Craig Roberts, Robert Shrock, Ivan Schmidt, Peter Tandy, and Xing-Gang Wu.
This research was supported by the Department of Energy,  contract DE--AC02--76SF00515.  
SLAC-PUB-16874.



\begin{thebibliography}{30}



\bibitem{Dirac:1949cp} 
  P.~A.~M.~Dirac,
  Rev.\ Mod.\ Phys.\  {\bf 21}, 392 (1949).
  doi:10.1103/RevModPhys.21.392


\bibitem{Brodsky:1997de} 
  S.~J.~Brodsky, H.~C.~Pauli and S.~S.~Pinsky,
  Phys.\ Rept.\  {\bf 301}, 299 (1998)
  doi:10.1016/S0370-1573(97)00089-6
  [hep-ph/9705477].


\bibitem{Brodsky:2000xy} 
  S.~J.~Brodsky, M.~Diehl and D.~S.~Hwang,
  Nucl.\ Phys.\ B {\bf 596}, 99 (2001)
  doi:10.1016/S0550-3213(00)00695-7
  [hep-ph/0009254].


\bibitem{Terrell:1959zz} 
  J.~Terrell,
  Phys.\ Rev.\  {\bf 116}, 1041 (1959).
  doi:10.1103/PhysRev.116.1041


\bibitem{Penrose:1959vz} 
  R.~Penrose,
  Proc.\ Cambridge Phil.\ Soc.\  {\bf 55}, 137 (1959).
  doi:10.1017/S0305004100033776



\bibitem{Brodsky:2000ii} 
  S.~J.~Brodsky, D.~S.~Hwang, B.~Q.~Ma and I.~Schmidt,
  Nucl.\ Phys.\ B {\bf 593}, 311 (2001)
  doi:10.1016/S0550-3213(00)00626-X
  [hep-th/0003082].


\bibitem{Kobzarev:1962wt} 
  I.~Y.~Kobzarev and L.~B.~Okun,
  Zh.\ Eksp.\ Teor.\ Fiz.\  {\bf 43}, 1904 (1962)
  [Sov.\ Phys.\ JETP {\bf 16}, 1343 (1963)].


\bibitem{Teryaev:1999su} 
  O.~V.~Teryaev,
  hep-ph/9904376.


\bibitem{Gribov:1972ri} 
  V.~N.~Gribov and L.~N.~Lipatov,
  Sov.\ J.\ Nucl.\ Phys.\  {\bf 15}, 438 (1972)
  [Yad.\ Fiz.\  {\bf 15}, 781 (1972)].


\bibitem{Altarelli:1977zs} 
  G.~Altarelli and G.~Parisi,
  Nucl.\ Phys.\ B {\bf 126}, 298 (1977).
  doi:10.1016/0550-3213(77)90384-4


\bibitem{Dokshitzer:1977sg} 
  Y.~L.~Dokshitzer,
  Sov.\ Phys.\ JETP {\bf 46}, 641 (1977)
  [Zh.\ Eksp.\ Teor.\ Fiz.\  {\bf 73}, 1216 (1977)].


\bibitem{Lepage:1979zb} 
  G.~P.~Lepage and S.~J.~Brodsky,
  Phys.\ Lett.\  {\bf 87B}, 359 (1979).
  doi:10.1016/0370-2693(79)90554-9


\bibitem{Lepage:1980fj} 
  G.~P.~Lepage and S.~J.~Brodsky,
  Phys.\ Rev.\ D {\bf 22}, 2157 (1980).
  doi:10.1103/PhysRevD.22.2157


\bibitem{Efremov:1979qk} 
  A.~V.~Efremov and A.~V.~Radyushkin,
  Phys.\ Lett.\  {\bf 94B}, 245 (1980).
  doi:10.1016/0370-2693(80)90869-2


\bibitem{Efremov:1978rn} 
  A.~V.~Efremov and A.~V.~Radyushkin,
  Theor.\ Math.\ Phys.\  {\bf 42}, 97 (1980)
  [Teor.\ Mat.\ Fiz.\  {\bf 42}, 147 (1980)].
  doi:10.1007/BF01032111


\bibitem{Brodsky:2015uwa} 
  S.~J.~Brodsky and S.~Gardner,
  Phys.\ Rev.\ Lett.\  {\bf 116}, no. 1, 019101 (2016)
  doi:10.1103/PhysRevLett.116.019101
  [arXiv:1504.00969 [hep-ph]].


\bibitem{Pauli:1985pv} 
  H.~C.~Pauli and S.~J.~Brodsky,
  Phys.\ Rev.\ D {\bf 32}, 1993 (1985).
  doi:10.1103/PhysRevD.32.1993


\bibitem{Hornbostel:1988fb} 
  K.~Hornbostel, S.~J.~Brodsky and H.~C.~Pauli,
  Phys.\ Rev.\ D {\bf 41}, 3814 (1990).
  doi:10.1103/PhysRevD.41.3814
  
\bibitem{deTeramond:2008ht} 
  G.~F.~de Teramond and S.~J.~Brodsky,
  Phys.\ Rev.\ Lett.\  {\bf 102}, 081601 (2009)
  doi:10.1103/PhysRevLett.102.081601
  [arXiv:0809.4899 [hep-ph]].


\bibitem{Vary:2014tqa} 
  J.~P.~Vary, X.~Zhao, A.~Ilderton, H.~Honkanen, P.~Maris and S.~J.~Brodsky,
  Nucl.\ Phys.\ Proc.\ Suppl.\  {\bf 251-252}, 10 (2014)
  doi:10.1016/j.nuclphysbps.2014.04.002
  [arXiv:1406.1838 [nucl-th]].


\bibitem{Brodsky:2015aia} 
  S.~J.~Brodsky {\it et al.},
  arXiv:1502.05728 [hep-ph].



\bibitem{deAlfaro:1976je}
 V.~de Alfaro, S.~Fubini and G.~Furlan,
  Nuovo Cim.\ A {\bf 34}, 569 (1976).

\bibitem{Brodsky:2013ar} 
  S.~J.~Brodsky, G.~F.~De TŽramond and H.~G.~Dosch,
  Phys.\ Lett.\ B {\bf 729}, 3 (2014)
  doi:10.1016/j.physletb.2013.12.044
  [arXiv:1302.4105 [hep-th]].





\bibitem{Brodsky:2009gx} 
  S.~J.~Brodsky and R.~F.~Lebed,
  Phys.\ Rev.\ Lett.\  {\bf 102}, 213401 (2009)
  doi:10.1103/PhysRevLett.102.213401
  [arXiv:0904.2225 [hep-ph]].


\bibitem{Banburski:2012tk} 
  A.~Banburski and P.~Schuster,
  Phys.\ Rev.\ D {\bf 86}, 093007 (2012)
  doi:10.1103/PhysRevD.86.093007
  [arXiv:1206.3961 [hep-ph]].


\bibitem{deTeramond:2013it} 
  G.~F.~de Teramond, H.~G.~Dosch and S.~J.~Brodsky,
  Phys.\ Rev.\ D {\bf 87}, no. 7, 075005 (2013)
  doi:10.1103/PhysRevD.87.075005
  [arXiv:1301.1651 [hep-ph]].


\bibitem{Brodsky:2014yha} 
  S.~J.~Brodsky, G.~F.~de Teramond, H.~G.~Dosch and J.~Erlich,
  Phys.\ Rept.\  {\bf 584}, 1 (2015)
  doi:10.1016/j.physrep.2015.05.001
  [arXiv:1407.8131 [hep-ph]].


\bibitem{Brodsky:2011xx} 
  S.~J.~Brodsky, F.~G.~Cao and G.~F.~de Teramond,
  Phys.\ Rev.\ D {\bf 84}, 075012 (2011)
  doi:10.1103/PhysRevD.84.075012
  [arXiv:1105.3999 [hep-ph]].


\bibitem{Forshaw:2012im} 
  J.~R.~Forshaw and R.~Sandapen,
  Phys.\ Rev.\ Lett.\  {\bf 109}, 081601 (2012)
  doi:10.1103/PhysRevLett.109.081601
  [arXiv:1203.6088 [hep-ph]].


\bibitem{Smirnov:2009fh} 
  A.~V.~Smirnov, V.~A.~Smirnov and M.~Steinhauser,
  Phys.\ Rev.\ Lett.\  {\bf 104}, 112002 (2010)
  doi:10.1103/PhysRevLett.104.112002
  [arXiv:0911.4742 [hep-ph]].


\bibitem{Haag:1974qh} 
  R.~Haag, J.~T.~Lopuszanski and M.~Sohnius,
  Nucl.\ Phys.\ B {\bf 88}, 257 (1975).
  doi:10.1016/0550-3213(75)90279-5


\bibitem{Fubini:1984hf} 
  S.~Fubini and E.~Rabinovici,
  Nucl.\ Phys.\ B {\bf 245}, 17 (1984).
  doi:10.1016/0550-3213(84)90422-X


\bibitem{deTeramond:2014asa} 
  G.~F.~de Teramond, H.~G.~Dosch and S.~J.~Brodsky,
  Phys.\ Rev.\ D {\bf 91}, no. 4, 045040 (2015)
  doi:10.1103/PhysRevD.91.045040
  [arXiv:1411.5243 [hep-ph]].


\bibitem{Dosch:2015nwa} 
  H.~G.~Dosch, G.~F.~de Teramond and S.~J.~Brodsky,
  Phys.\ Rev.\ D {\bf 91}, no. 8, 085016 (2015)
  doi:10.1103/PhysRevD.91.085016
  [arXiv:1501.00959 [hep-th]].


\bibitem{Dosch:2015bca} 
  H.~G.~Dosch, G.~F.~de Teramond and S.~J.~Brodsky,
  Phys.\ Rev.\ D {\bf 92}, no. 7, 074010 (2015)
  doi:10.1103/PhysRevD.92.074010
  [arXiv:1504.05112 [hep-ph]].


\bibitem{Liu:2015jna} 
  T.~Liu and B.~Q.~Ma,
  Phys.\ Rev.\ D {\bf 92}, no. 9, 096003 (2015)
  doi:10.1103/PhysRevD.92.096003
  [arXiv:1510.07783 [hep-ph]].


\bibitem{Brodsky:2014hia} 
  S.~J.~Brodsky,
  Nucl.\ Part.\ Phys.\ Proc.\  {\bf 258-259}, 23 (2015)
  doi:10.1016/j.nuclphysbps.2015.01.007
  [arXiv:1410.0404 [hep-ph]].


\bibitem{Bjorken:2013boa} 
  J.~D.~Bjorken, S.~J.~Brodsky and A.~Scharff Goldhaber,
  Phys.\ Lett.\ B {\bf 726}, 344 (2013)
  doi:10.1016/j.physletb.2013.08.066
  [arXiv:1308.1435 [hep-ph]].


\bibitem{Cruz-Santiago:2015dla} 
  C.~Cruz-Santiago, P.~Kotko and A.~M.~Sta?to,
  Prog.\ Part.\ Nucl.\ Phys.\  {\bf 85}, 82 (2015).
  doi:10.1016/j.ppnp.2015.07.002
  
  
\bibitem{Kelly} 
K. Chiu and S. J. Brodsky, to be published.

\bibitem{Brodsky:1973kb} 
  S.~J.~Brodsky, R.~Roskies and R.~Suaya,
  Phys.\ Rev.\ D {\bf 8}, 4574 (1973).
  doi:10.1103/PhysRevD.8.4574


\bibitem{Brodsky:2009dr} 
  S.~J.~Brodsky and G.~F.~de Teramond,
  arXiv:0901.0770 [hep-ph].
  
  \bibitem{Ashery:2000yj} 
  D.~Ashery,
  ``Measurement of light cone wave functions by diffractive dissociation,''
  Nucl.\ Phys.\ Proc.\ Suppl.\  {\bf 90}, 67 (2000)
doi:10.1016/S0920-5632(00)00875-6, 10.1016/S0920-5632(02)01354-3
  [hep-ex/0008036].



\bibitem{Munger:1993kq} 
  C.~T.~Munger, S.~J.~Brodsky and I.~Schmidt,
  Phys.\ Rev.\ D {\bf 49}, 3228 (1994).
  doi:10.1103/PhysRevD.49.3228


\bibitem{Zee:2008zz} 
  A.~Zee,
  Mod.\ Phys.\ Lett.\ A {\bf 23}, 1336 (2008).
  doi:10.1142/S0217732308027709



\bibitem{Brodsky:2010xf} 
  S.~J.~Brodsky, C.~D.~Roberts, R.~Shrock and P.~C.~Tandy,
  Phys.\ Rev.\ C {\bf 82}, 022201 (2010)
  doi:10.1103/PhysRevC.82.022201
  [arXiv:1005.4610 [nucl-th]].


\bibitem{Casher:1974xd} 
  A.~Casher and L.~Susskind,
  Phys.\ Rev.\ D {\bf 9}, 436 (1974).
  doi:10.1103/PhysRevD.9.436


\bibitem{Brodsky:2009zd} 
  S.~J.~Brodsky and R.~Shrock,
  Proc.\ Nat.\ Acad.\ Sci.\  {\bf 108}, 45 (2011)
  doi:10.1073/pnas.1010113107
  [arXiv:0905.1151 [hep-th]].


\bibitem{Srivastava:2002mw} 
  P.~P.~Srivastava and S.~J.~Brodsky,
  Phys.\ Rev.\ D {\bf 66}, 045019 (2002)
  doi:10.1103/PhysRevD.66.045019
  [hep-ph/0202141].


\bibitem{Verlinde:2016toy} 
  E.~P.~Verlinde,
  arXiv:1611.02269 [hep-th].


\bibitem{Grunberg:1980ja} 
  G.~Grunberg,
  Phys.\ Lett.\  {\bf 95B}, 70 (1980)
  Erratum: [Phys.\ Lett.\  {\bf 110B}, 501 (1982)].
  doi:10.1016/0370-2693(80)90402-5


\bibitem{Brodsky:1994eh} 
  S.~J.~Brodsky and H.~J.~Lu,
  Phys.\ Rev.\ D {\bf 51}, 3652 (1995)
  doi:10.1103/PhysRevD.51.3652
  [hep-ph/9405218].


\bibitem{Brodsky:2010ur} 
  S.~J.~Brodsky, G.~F.~de Teramond and A.~Deur,
  Phys.\ Rev.\ D {\bf 81}, 096010 (2010)
  doi:10.1103/PhysRevD.81.096010
  [arXiv:1002.3948 [hep-ph]].


\bibitem{Deur:2005cf} 
  A.~Deur, V.~Burkert, J.~P.~Chen and W.~Korsch,
  Phys.\ Lett.\ B {\bf 650}, 244 (2007)
  doi:10.1016/j.physletb.2007.05.015
  [hep-ph/0509113].


\bibitem{Deur:2014qfa} 
  A.~Deur, S.~J.~Brodsky and G.~F.~de Teramond,
  Phys.\ Lett.\ B {\bf 750}, 528 (2015)
  doi:10.1016/j.physletb.2015.09.063
  [arXiv:1409.5488 [hep-ph]].


\bibitem{Brodsky:2014jia} 
  S.~J.~Brodsky, G.~F.~de TŽramond, A.~Deur and H.~G.~Dosch,
  Few Body Syst.\  {\bf 56}, no. 6-9, 621 (2015)
  doi:10.1007/s00601-015-0964-1
  [arXiv:1410.0425 [hep-ph]].


\bibitem{Agashe:2014kda} 
  K.~A.~Olive {\it et al.} [Particle Data Group Collaboration],
  Chin.\ Phys.\ C {\bf 38}, 090001 (2014).
  doi:10.1088/1674-1137/38/9/090001


\bibitem{Zee:2003mt} 
  A.~Zee,
  Princeton, UK: Princeton Univ. Pr. (2010) 576 p




\bibitem{Mojaza:2012mf} 
  M.~Mojaza, S.~J.~Brodsky and X.~G.~Wu,
  Phys.\ Rev.\ Lett.\  {\bf 110}, 192001 (2013)
  doi:10.1103/PhysRevLett.110.192001
  [arXiv:1212.0049 [hep-ph]].


\bibitem{Brodsky:1980zm} 
  S.~J.~Brodsky and S.~D.~Drell,
  Phys.\ Rev.\ D {\bf 22}, 2236 (1980).
  doi:10.1103/PhysRevD.22.2236


\bibitem{Liuti:2013cna} 
  S.~Liuti, A.~Rajan, A.~Courtoy, G.~R.~Goldstein and J.~O.~Gonzalez Hernandez,
  Int.\ J.\ Mod.\ Phys.\ Conf.\ Ser.\  {\bf 25}, 1460009 (2014)
  doi:10.1142/S201019451460009X
  [arXiv:1309.7029 [hep-ph]].


\bibitem{Mondal:2015uha} 
  C.~Mondal and D.~Chakrabarti,
  Eur.\ Phys.\ J.\ C {\bf 75}, no. 6, 261 (2015)
  doi:10.1140/epjc/s10052-015-3486-6
  [arXiv:1501.05489 [hep-ph]].


\bibitem{Lorce:2011dv} 
  C.~Lorce, B.~Pasquini and M.~Vanderhaeghen,
  JHEP {\bf 1105}, 041 (2011)
  doi:10.1007/JHEP05(2011)041
  [arXiv:1102.4704 [hep-ph]].


\bibitem{Brodsky:2008xe} 
  S.~J.~Brodsky,
  AIP Conf.\ Proc.\  {\bf 1105}, 315 (2009)
  doi:10.1063/1.3122202
  [arXiv:0811.0875 [hep-ph]].


\bibitem{Brodsky:2009dv} 
  S.~J.~Brodsky,
  Nucl.\ Phys.\ A {\bf 827}, 327C (2009)
  doi:10.1016/j.nuclphysa.2009.05.068
  [arXiv:0901.0781 [hep-ph]].


\bibitem{Brodsky:2002cx} 
  S.~J.~Brodsky, D.~S.~Hwang and I.~Schmidt,
  Phys.\ Lett.\ B {\bf 530}, 99 (2002)
  doi:10.1016/S0370-2693(02)01320-5
  [hep-ph/0201296].


\bibitem{Brodsky:2002ue} 
  S.~J.~Brodsky, P.~Hoyer, N.~Marchal, S.~Peigne and F.~Sannino,
  Phys.\ Rev.\ D {\bf 65}, 114025 (2002)
  doi:10.1103/PhysRevD.65.114025
  [hep-ph/0104291].


\bibitem{Brodsky:2010vs} 
  S.~J.~Brodsky, B.~Pasquini, B.~W.~Xiao and F.~Yuan,
  Phys.\ Lett.\ B {\bf 687}, 327 (2010)
  doi:10.1016/j.physletb.2010.03.049
  [arXiv:1001.1163 [hep-ph]].


\bibitem{Brodsky:2013oya} 
  S.~J.~Brodsky, D.~S.~Hwang, Y.~V.~Kovchegov, I.~Schmidt and M.~D.~Sievert,
  Phys.\ Rev.\ D {\bf 88}, no. 1, 014032 (2013)
  doi:10.1103/PhysRevD.88.014032
  [arXiv:1304.5237 [hep-ph]].


\bibitem{Brodsky:1989qz} 
  S.~J.~Brodsky and H.~J.~Lu,
  Phys.\ Rev.\ Lett.\  {\bf 64}, 1342 (1990).
  doi:10.1103/PhysRevLett.64.1342


\bibitem{Brodsky:2004qa} 
  S.~J.~Brodsky, I.~Schmidt and J.~J.~Yang,
  Phys.\ Rev.\ D {\bf 70}, 116003 (2004)
  doi:10.1103/PhysRevD.70.116003
  [hep-ph/0409279].


\bibitem{Schienbein:2007fs} 
  I.~Schienbein, J.~Y.~Yu, C.~Keppel, J.~G.~Morfin, F.~Olness and J.~F.~Owens,
  Phys.\ Rev.\ D {\bf 77}, 054013 (2008)
  doi:10.1103/PhysRevD.77.054013
  [arXiv:0710.4897 [hep-ph]].




 
\end{thebibliography}
\end{document}